\documentclass[aps,twocolumn,superscriptaddress,showpacs,longbibliography]{revtex4-1}

\usepackage{graphicx}
\usepackage{dcolumn}
\usepackage{bm}
\usepackage{upgreek}
\usepackage{paralist}
\usepackage{gensymb}
\usepackage{braket}

\newcommand{\EECS}{\affiliation{Research Laboratory of Electronics, Massachusetts Institute of Technology, Cambridge, MA 02139, USA}}
\newcommand{\Basel}{\affiliation{Department of Physics, University of Basel, Klingelbergstrasse 82, CH-4056 Basel, Switzerland}}
\newcommand{\Copenhagen}{\affiliation{Niels Bohr Institute, University of Copenhagen, 2100 Copenhagen, Denmark}}
\newcommand{\Albany}{\affiliation{College of Nanoscale Science and Engineering, SUNY Polytechnic Institute, Albany, NY 12203, USA}}
\begin{document}

\title{Efficient Extraction of Light from a Nitrogen-Vacancy Center in a Diamond Parabolic Reflector}

\author{Noel H. Wan}
\email{noelwan@mit.edu}
\EECS
\author{Brendan J. Shields}
\Basel
\author{Donggyu Kim}
\EECS
\author{Sara Mouradian}
\EECS
\author{Benjamin Lienhard}
\EECS
\author{Michael Walsh}
\EECS
\author{Hassaram Bakhru}
\Albany
\author{Tim Schr{\"o}der}
\EECS
\Copenhagen
\author{Dirk Englund}
\email{englund@mit.edu}
\EECS

\begin{abstract}

Quantum emitters in solids are being developed for a range of quantum technologies, including quantum networks,  computing, and sensing. However, a remaining challenge is the poor photon collection due to the high refractive index of most host materials. Here we overcome this limitation by introducing monolithic parabolic reflectors as an efficient geometry for broadband photon extraction from quantum emitter and experimentally demonstrate this device for the nitrogen-vacancy (NV) center in diamond. Simulations indicate a photon collection efficiency exceeding 75\% across the visible spectrum and experimental devices, fabricated using a high-throughput gray-scale lithography process, demonstrate a photon extraction efficiency of $(48\pm 5)$\%. This device enables a raw experimental efficiency of $(12\pm 2)$\% with fluorescence detection rates as high as $(4.6 - 5.7)\times 10^6$ counts per second from a single NV center. 

\end{abstract}
\maketitle

\begin{figure*}[ht]
\begin{center}
\includegraphics[width=6in]
{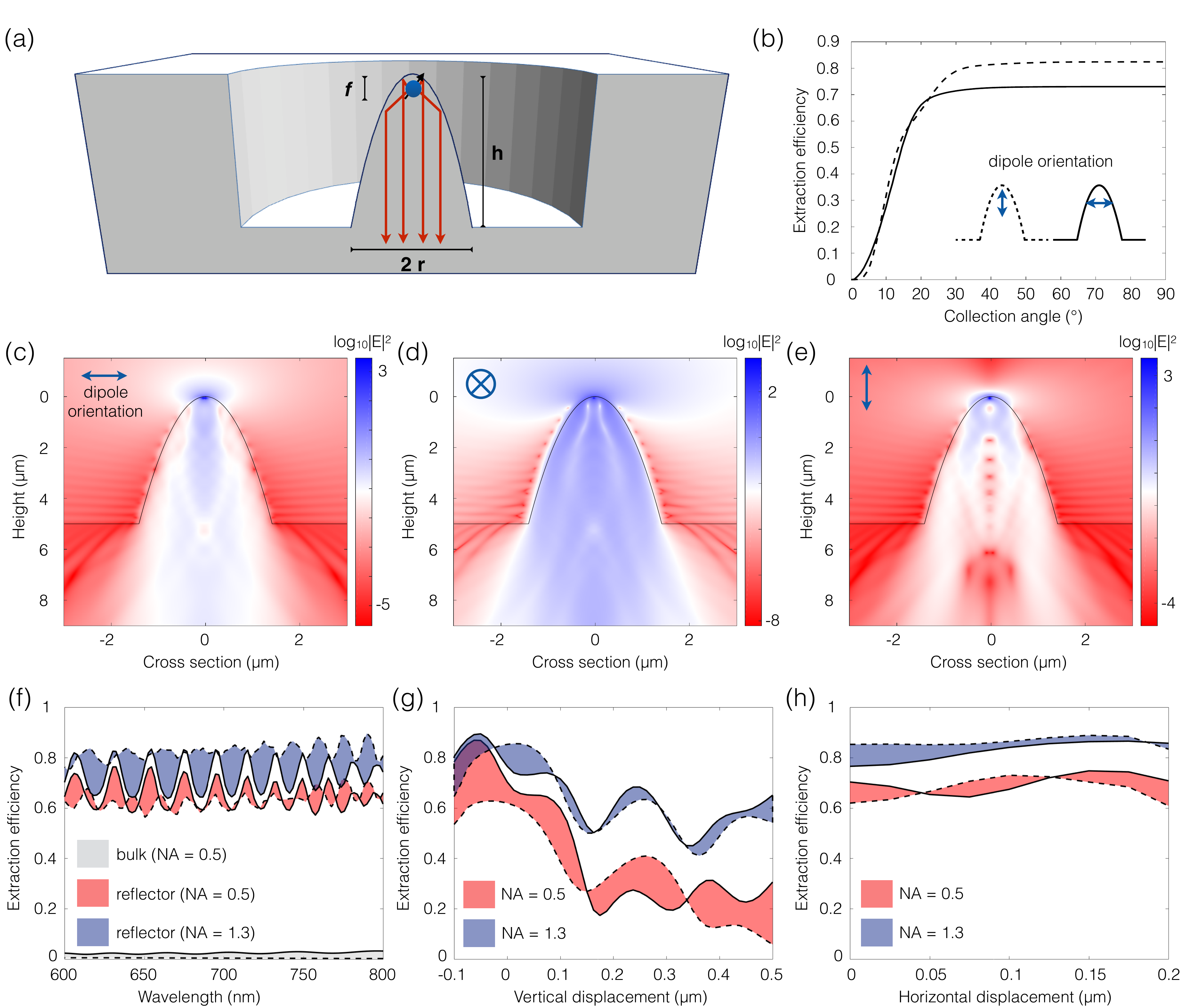}
\end{center}
\caption{(a) Geometrical optics illustration of the photonic device with a single NV center located at its focus $f$. (b) Mean collection efficiency for emission wavelengths $\lambda = 600 - 800$\,nm as a function of collection angle for a dipole oriented perpendicular and parallel to the axis of the parabolic reflector. (c-e) Finite-difference time-domain simulations of a device with $f = 100$\,nm and height $h = 5$\,$\upmu$m for a dipole oriented perpendicular in-plane, perpendicular out-of-plane and parallel to the paraboloid axis, respectively. (f) Simulated collection efficiency for NA\,=\,0.5 and NA\,=\,1.3 for a dipole oriented perpendicularly (solid line) and in parallel (dashed line), respectively. For comparison, shown in gray is the collection efficiency of an NV located in an unpatterned diamond ($\eta \approx 0.05$). (g) Collection efficiency at 637\,nm as a function of a vertical displacement of the emitter. (h) Collection efficiency at 637\,nm as a function of a horizontal displacement of the emitter.
}
\label{concept}
\end{figure*}

Solid-state single-photon emitters and spin qubits offer unique advantages as scalable building blocks for emerging quantum technologies~\cite{aharonovich2016solid}. Numerous demonstrations have established these platforms as attractive testbeds for investigating fundamental physics~\cite{hensen2015loophole} and quantum optical phenomena~\cite{buckley2012engineered,lodahl2015interfacing}, and implementing quantum networks~\cite{togan2010quantum,de2012quantum,gao2012observation,bernien2013heralded,pfaff2014unconditional,delteil2016generation,kalb2017entanglement,stockill2017phase}. Scaling these systems would require new solutions to existing bottlenecks, chiefly the low photon collection efficiency, which arises due to substantial total internal reflection in solids. As single-photon source, a low photon count rate directly translates to a small detection probability, precluding the efficient generation of large entangled states. As spin qubits, poor optical readout of spin states limits the number of operations within the quantum memory lifetime. Through quantum engineering, some of these platforms are now rapidly maturing as promising practical technologies. For example, near-unity single-photon sources have been realized with single molecules coupled to dielectric antennas~\cite{chu2016single} and quantum dots (QDs) embedded in a photonic crystal waveguide~\cite{arcari2014near}, and QD-microcavity systems~\cite{gazzano2013bright,somaschi2016near,ding2016demand}  have been recently operated as high rate, high purity indistinguishable single-photon sources for boson sampling~\cite{wang2017high}. In this work, we address the photon collection problem for the nitrogen-vacancy (NV) spin qubit in diamond, although the concept also works for any emitter in a high index material. 

The negatively charged nitrogen-vacancy color center is a substitutional defect in diamond with atom-like optical transitions and millisecond-long spin coherence times~\cite{doherty2013nitrogen}. It is an exceptionally versatile qubit~\cite{childress2014atom} with applications ranging from nanoscale quantum sensing~\cite{rondin2014magnetometry} to quantum information processing~\cite{childress2013diamond}. High fidelity quantum control of the NV electron and nuclear spins, along with demonstrations of heralded spin-spin entanglement have established the NV as a leading system in a quantum node. Central to the operation of the NV qubit is its spin-dependent fluorescence -- an NV center appears bright or dark depending on its spin state. Improving the efficiency of this spin readout presents a key challenge due to the high refractive index of diamond, which results in strong confinement via total internal reflection. As such, there is an intense research effort into building highly efficient optical interfaces in diamond~\cite{lonvcar2013quantum,aharonovich2014diamond,schroder2016quantum}. 
In addition to the required high photon extraction efficiency, different applications introduce varying technical demands on the NV-optical interface. To entangle distant NV centers via quantum interference, the photon wavepackets should be indistinguishable in frequency and maintain excellent optical coherence~\cite{bernien2013heralded,pfaff2014unconditional,hensen2015loophole,kalb2017entanglement}. Whereas the former can be controlled via the dc Stark effect~\cite{tamarat2006stark,bassett2011electrical}, the latter demonstrations have mainly relied on NV centers that are natural~\cite{tamarat2006stark}, or deep in the bulk crystal~\cite{chen2017laser} or in a solid immersion lens~\cite{bernien2012two,bernien2013heralded,pfaff2014unconditional,hensen2015loophole} for stable optical transitions. Recently, NV centers with coherent transitions have been realized by nitrogen implantation $\sim$ 100\,nm below the diamond surface~\cite{chu2014coherent}, opening the prospects for scalable integration of NV qubits with photonic devices for enhanced collection efficiency and light-matter interactions. Within this paradigm, one of the technical challenges is the preservation of the NV's nearly transform-limited lines, which so far has not been observed in diamond nanophotonic devices~\cite{faraon2012coupling,li2015coherent,riedel2017deterministic}. As with the case of a solid immersion lens, one reasonable route towards optically stable systems is therefore through the coupling to devices that place surfaces at least 100\,nm from the NV. 

While NV centers as quantum memories can be optically connected via the interference of the narrow zero-phonon-line (ZPL) emission, NVs as quantum sensors take full advantage of both the coherent ZPL and the phonon-sideband (PSB) emission, which span over $\sim 150$\, nm in wavelength. This condition favors non-resonant and spectrally broadband devices. In addition, in many sensing applications such as nanoscale magnetic resonance imaging and scanning tip magnetometry~\cite{balasubramanian2008nanoscale,maletinsky2012robust,rondin2013stray,appel2016fabrication,pelliccione2016scanned} the optical interface should provide direct access between the NV probe sensor and the target. Therefore, there is also a demand for compatibility with near-surface ($<30$\,nm) NV centers.

The diamond parabolic reflector introduced here satisfies the various requirements listed above. Our device consists of a micro-reflector fabricated in bulk diamond around a quantum emitter, as shown in Fig.~\ref{concept}a. The design is based on geometrical optics considerations, and is inherently broadband and compatible with various NV depths. To produce this three-dimensional single-photon source device, we developed a gray-scale fabrication process using standard electron-beam lithography and reactive-ion etching techniques. To the best of our knowledge, our first experiments here represent the brightest single-photon source in diamond, with photon extraction efficiency exceeding 48\% and fluorescence rates greater than 4.6\,MHz. As this device relies only on the index contrast with air, the design also readily applies to solid-state quantum emitters in a variety of host materials, including other optically active defects in diamond~\cite{zaitsev2013optical}, silicon carbide~\cite{lohrmann2017review}, and III-nitride compounds~\cite{aharonovich2016solid}, as well as quantum dots in III-V heterostructures~\cite{buckley2012engineered}.

\begin{figure*}[ht]
\begin{center}
\includegraphics[width=6in]
{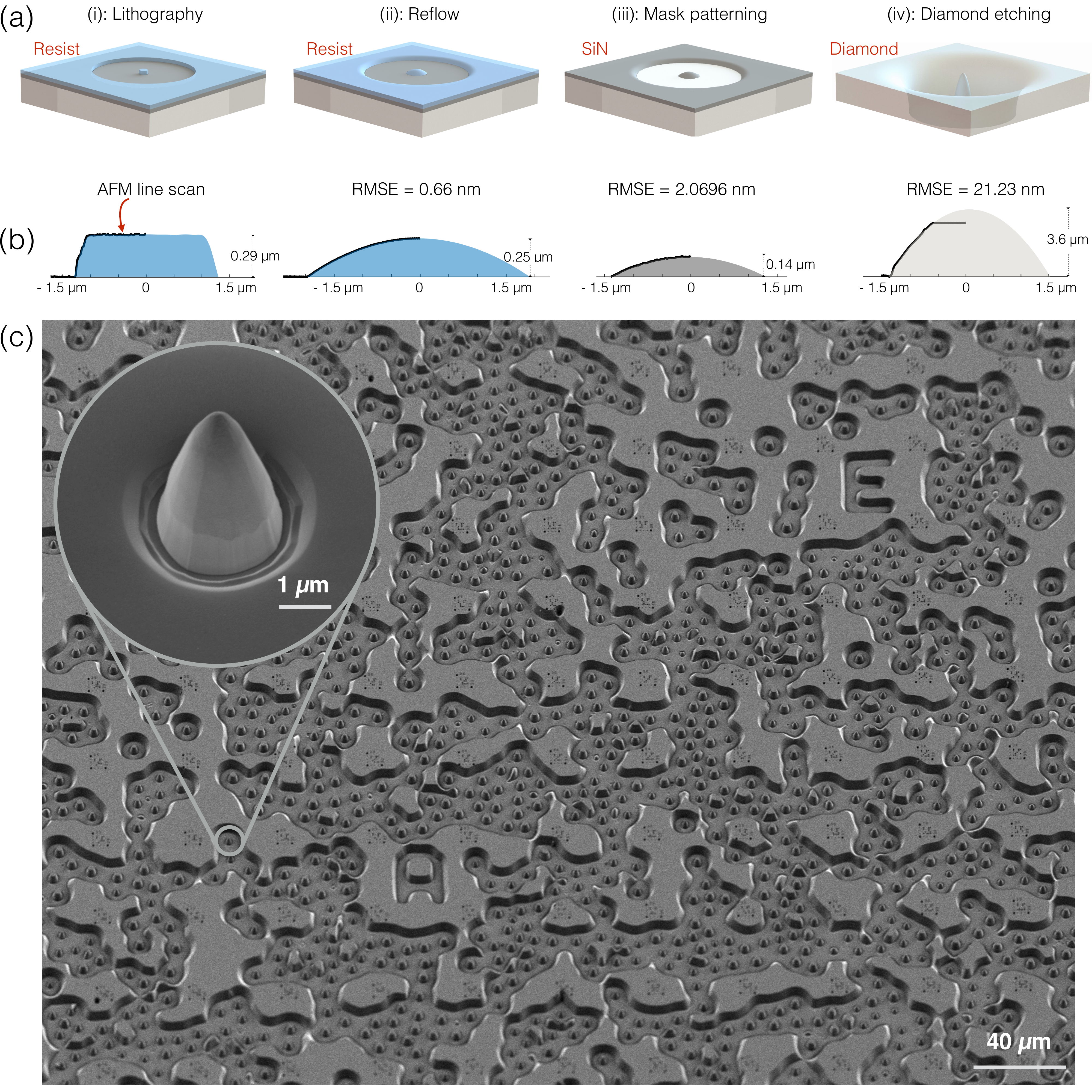}
\end{center}
\caption{(a) Schematic of fabrication steps. (b) Atomic-force microscopy (AFM) cross section linescans of a device during the corresponding fabrication step. RMSE denotes the root-mean-squared-error of the AFM linescan from the underlying model (see Methods). (i) Electron-beam lithography defines a disk of resist (blue) over a SiN (gray) and an NV center. (ii) Thermal reflow produces near-hemispherical shape as shown by the AFM linescan (dashed lines). (iii) Reactive ion etching transfers this profile into SiN. (iv) The SiN serves as a grey-scale etch mask for diamond. The desired parabola is produced because the dry etch rate into SiN is $\sim 28$ times slower than into diamond. See main text and Methods for more details regarding the fabrication. (c) Scanning electron micrograph of the diamond chip in this experiment. The devices on this chip have a height of 5\,$\upmu$m. The inset shows an example of a produced parabolic reflector. The irregular arrangement of devices on the diamond is due to a lithography process that registers the masks to pre-localized NV centers.
}
\label{fab}
\end{figure*}

The high refractive index ($n = 2.4$) of diamond permits only light with a narrow emission angle ($\theta_{\text{crit}} \approx 24\degree$) to escape the bulk material, while higher angles are reflected by total internal reflection (TIR). In many diamond NV experiments~\cite{hadden2010strongly,marseglia2011nanofabricated,schroder2011ultrabright,jamali2014microscopic,waldherr2014quantum,sipahigil2012quantum,bernien2012two,bernien2013heralded,pfaff2014unconditional,hensen2015loophole,kalb2017entanglement}, a solid immersion lens is typically milled around an NV center to reduce TIR in one direction. Rather than avoiding TIR, our parabolic reflector exploits TIR to reflect emission downward for collection from the opposite diamond surface. In the numerical design of the structure, we consider a dipole emitter located at the focus ($f$) of a paraboloidal diamond structure, as illustrated in Fig.\,\ref{concept}a. We verify this geometrical optics argument using finite-difference time-domain (FDTD) simulations. Figure.\,\ref{concept}b demonstrates high directionality of the collected photons, approaching its maximum around 30$\degree$, which corresponds to a numerical aperture (NA) of 0.5. We show the intensity distribution in Fig.\,\ref{concept}c-e for a dipole oriented orthogonal or parallel to the axis of the paraboloid. In these simulations, as in our experiments, we consider a device height of $h = 5$\,$\upmu$m with a single NV implanted at the depth of $f = 100$ nm. As shown in Fig.\,\ref{concept}f, $\sim 77$\% of emitted light is collected into an NA of 1.3, a $\sim$20-fold improvement in efficiency over collection from an unpatterned diamond. This structure is also robust to fabrication imperfections, where deviations from an ideal curvature result in displaced foci, or equivalently, a displaced emitter. Simulations indicate that the efficiency remains above 60\% for when the emitter is vertically displaced by 200\,nm (Fig.\,\ref{concept}g). Similarly, the structure is insensitive to small lateral displacements of up to 200\,nm (Fig.\,\ref{concept}h)

\begin{figure*}[ht]
\begin{center}
\includegraphics[width=7in]
{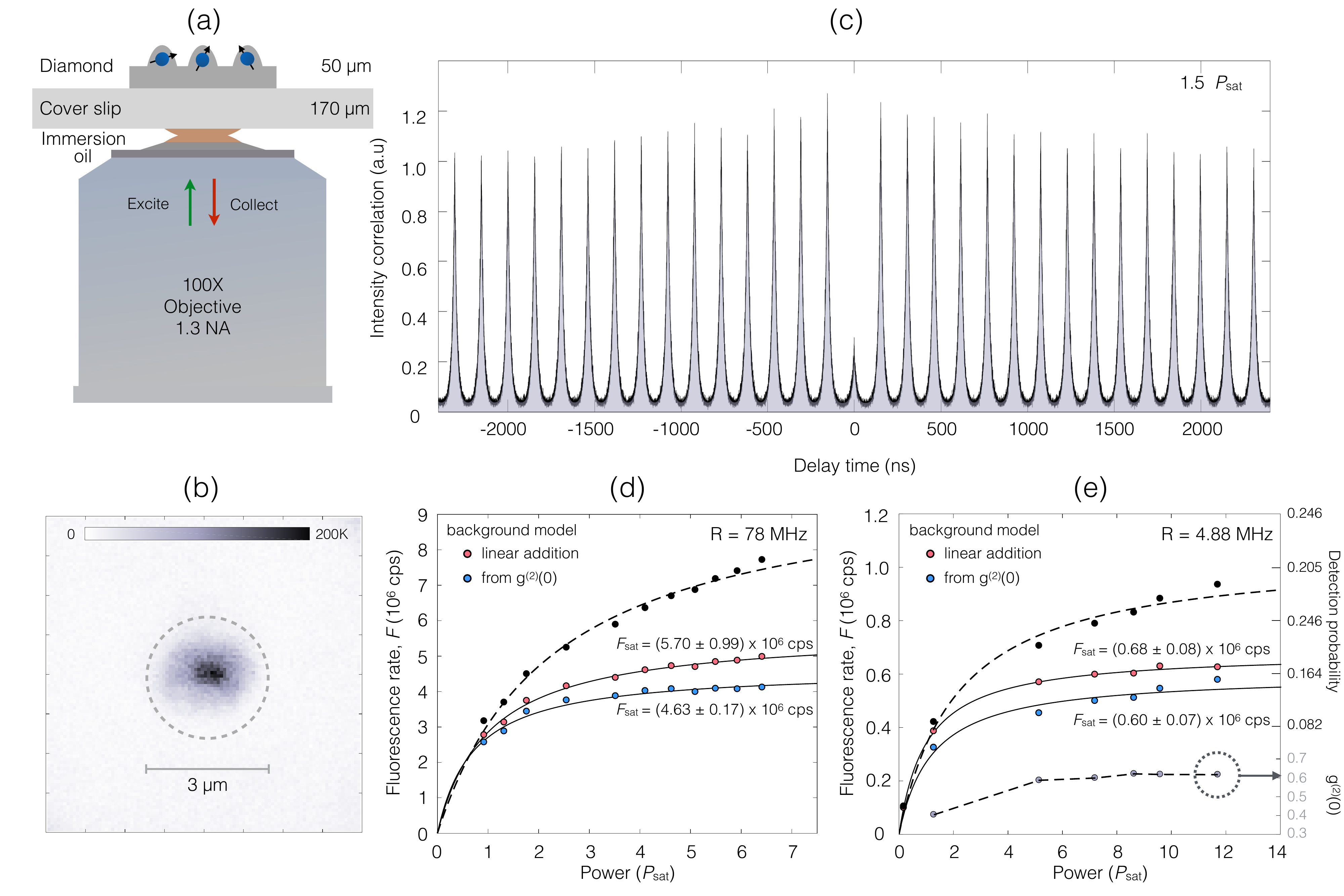}
\end{center}
\caption{(a) Experimental configuration using an inverted confocal microscope. We collect and redirect the NV fluorescence for photon count rate measurement of Hanbury Brown-Twiss interferometry. (b) Scanning confocal photoluminescence (PL) image of a single NV center. (c) The second-order correlation function at zero-delay of the PL is $g^{(2)}(0) = 0.1873(5)$ at $P = 1.5 P_{\text{sat}}$, indicating the presence of a single emitter ($g^{(2)}(0)<0.5$). (d) Power-dependent fluorescence count rate measurements indicate a maximum $F_{\text{sat}} = (5.70\pm 0.99) \times 10^6$ counts per second after background correction\,(red). Using another background correction method (see main text), the maximum count rate is $F_{\text{sat}} = (4.63 \pm 0.17) \times 10^6$ counts per second\,(blue). In this experiment, the NV is excited every $\tau_{\text{rad}}$, or at a rate of 78\,MHz, and $P_{\text{sat}} = 0.32$\,mW.  (e) By reducing the repetition rate to 4.88\,MHz which corresponds to approximately one excitation pulse every $16\tau_{\text{rad}}$, we obtained a detection probability of $0.14 \pm 0.01$\,(red) and $0.12 \pm 0.01$\,(blue) using the two methods outlined in the text. Here, $P_{\text{sat}} = 0.025$\,mW, and from these measurements, we determine a brightness, defined as the fraction of extracted photons per spontaneous emission, of at least $(0.48\pm 5$) from this NV center. 
} 
\label{results}
\end{figure*}

To fabricate these parabolic reflectors in diamond, we developed a novel reactive-ion etching technique using an atomically smooth gray-scale hardmask, Fig.2. The processing begins by depositing a $\sim$ 200-nm thick layer of silicon nitride (SiN) on a diamond chip. Electron-beam lithography defines disks of a 280-nm thick ZEP 520A polymer electron beam resist (EBR) over pre-localized NV centers (Fig.\,\ref{fab}a(i)). Heating the sample above the EBR glass transition temperature causes it to reflow into a curved surface shown in Fig.\,\ref{fab}a(ii). We found experimentally that a temperature of 215$^{\circ}$C produces the required curvature. The atomic-force microscopy (AFM) in Fig.\,\ref{fab}b(ii) indicates an atomically smooth profile of the reflowed EBR. To replicate this profile in our SiN hard mask, we use an etch chemistry that faithfully transfers the pattern with an etch selectivity of $\sim$ 1, as indicated in Fig.\,\ref{fab}a(iii) and Fig.\,\ref{fab}b(iii). An oxygen plasma etch transfers this pattern into the diamond. This step exploits the oxygen etch selectivity of $\sim$28:1 between diamond and SiN to produce a diamond surface with elliptical profile (Fig.\,\ref{fab}a(iv)). With sufficient etching ($\sim 5\upmu$m), the structure well approximates a parabolic reflector. We verify a fabricated structure by AFM as shown in Fig.\,\ref{fab}b(iv), this time fitted to a parabolic surface (see Methods). Using this process combined with aligned lithography, we were able to fabricate parabolic reflectors over registered NV centers in diamond. Figure \,\ref{fab}c shows a scanning electron micrograph of the 50-$\upmu$m-thick diamond chip that was investigated in this work. 

We characterized the fabricated devices in an inverted confocal microscope with a non-resonant pulsed ($<10$ ps) 532 nm laser source and 635\,nm longpass filter  (Fig.\,\ref{results}a). The scanning confocal image in Fig.\,\ref{results}b shows an NV center coupled to a parabolic reflector. This system showed a normalized second-order correlation function at zero-delay of $g^{(2)}(0) = 0.1873(5)$ at $P = 1.5 P_{\text{sat}}$, as shown in Fig.\,\ref{results}c, measured using Hanbury Brown and Twiss interferometry with single photon avalanche diodes. The autocorrelation at zero-delay remains significantly below 0.5 with $g^{(2)}(0) = 0.4042(4)$ at $P = 1.3 P_{\text{sat}}$, even after increasing the aperture diameter of the confocal filter to $20\times$ the diffraction limit for better collection, demonstrating the low intrinsic background contribution from the system. 

Figure\,\ref{results}d shows the power-dependent fluorescence detection rate, $F$, when excited at the approximate inverse lifetime rate of $78.1$\,MHz. The black curve indicates the detected fluorescence rate, which includes contributions from single-photon event rate, $S$, and background and detector dark count rate, $B$. One way to quantify $B$ is to perform a control experiment without an emitter, which is accurate only for isotropic material environments. In situations where this is not true -- e.g. in patterned materials -- it is common to directly extract a linear background term in the detected fluorescence~\cite{schroder2012nanodiamond,momenzadeh2014nanoengineered,riedel2014low,li2015efficient,patel2016efficient,rodiek2017experimental} through
\begin{equation}\label{saturationcurve}
F(P) = \frac{F_{\text{sat}}}{1+P_{\text{sat}/P}} + \text{const}\cdot P
\end{equation} The first term expresses the emitter's saturable response and the second term represents the (unsaturable) linear response of the background photons. By subtracting this background term from the detected fluorescence, we arrive at $F_{\text{sat}} = (5.70\pm 0.99) \times 10^6$ counts per second, as shown by the red curve in Fig.\,\ref{results}d. 

To deduce the NV fluorescence using another method, we use the $g^{(2)}(0)$, which contains information about the photon purity in the coincidence histogram. At non-zero time delay, the coincidence histogram area is proportional to $(S+B)^2$; but at zero time delay, the area is proportional to $2SB + B^2$ because the NV cannot emit more than one photon per triggering~\cite{lounis2000single}. The normalized 
\begin{equation}
g^{2}(0) = 1-\frac{S^2}{(S+B)^2} 
\end{equation} therefore allows the extraction of $B$. The blue curve in Fig.\,\ref{results}d plots the fluorescence rate using this independent method, which reaches $F_{\text{sat}} = (4.63\pm 0.17) \times 10^6$ counts per second and a saturation power of $P_{\text{sat}} = 0.32$\,mW. Although the more commonly used saturation-curve estimation produces a higher fluorescence rate than the autocorrelation method (observed also in Ref.\,\cite{li2015efficient}), as far as we know, this lower bound is already the highest detected fluorescence count rate from a single NV center in diamond. 

While $F$ is frequently used as a metric of efficiency, practically it is highly sensitive to the emitter lifetime, electromagnetic environment, setup transmission, and detector efficiency. In addition, the ground state population is not unity at the onset of each excitation because the NV electronic state could still be in the excited triplet state, in the metastable single state or in a charge state other than the NV$^{-}$. For a more relevant comparison, we measured the NV$^{-}$ brightness $\eta_0$, defined here as the fraction of extracted photons per spontaneous emission~\cite{gazzano2013bright,somaschi2016near}. The power-dependent brightness can be expressed as
\begin{equation} \label{brightness}
\eta(P) = \eta_0\sigma(P),
\end{equation} where 
\begin{equation}\label{excitation_prob}
\sigma(P) = \frac{P/P_{\text{sat}}}{1+P/P_{\text{sat}}}(1-e^{-\tau_{\text{p}}/\tau_{\text{rad}}[1+P/P_{\text{sat}}]})
\end{equation} is the excited state population, $P_{\text{sat}}$  the saturation power, $\tau_{\text{rad}}$ the excited state radiative lifetime, and $\tau_{\text{p}}$ the pulse length~\cite{treussart2002direct}. 

Experimentally, the brightness is reduced by setup losses ($\eta_{\text{setup}}\approx 0.2538$), yielding instead a measurement of the triggering and detection probability. To ensure maximal ground state population for efficient single-photon generation, it is important to excite the NV at a rate that is significantly slower than the excited state population decay rate. From the shape of the correlation peaks in Fig.\,\ref{results}c, we extract the NV$^{-}$ radiative lifetime of $\tau_{\text{rad}} = 12.67$\,ns. We therefore used a pulse picker to lower the pulsed laser repetition rate to $R = 4.88$ MHz, which corresponds to one excitation pulse every $16\tau_{\text{rad}}$. Figure\,\ref{results}(e) reports the detected fluorescence (left axis) and single-photon triggering and detection probability (right axis). Again, using the two background correction methods as discussed earlier, we determine a maximum count rate of $(0.68\pm 0.11) \times 10^6$ (red curve) and $(0.60 \pm 0.07) \times 10^6$ (blue curve) with a saturation power of $P_{\text{sat}} = 0.025$\,mW, using Eq.\,(\ref{brightness}). These rates correspond to a probability of single-photon triggering and detection that is greater than ($0.12\pm 2$), which yields a brightness of at least $\eta_0 = 0.48\pm 0.05$. We also observe a similar brightness of $\eta_0 = 0.46\pm 0.04$ from another device. 

The efficient light collection presented here is promising for a variety of applications, in particular for quantum sensing and quantum networking based on atomic defects in solids. We note that heralded photon-based entanglement schemes with NV centers require additional characterizations that are beyond the scope of this work. In particular, cryogenic investigations and two-photon quantum interference experiments could reveal the indistinguishability and optical coherence of the emitted wavepackets. In addition to these experiments, several future modifications are expected to yield immediate, improved device performance. As the NV center undergoes photo-induced charge state switching between the optically detected NV$^{-}$ and the filtered NV$^{0}$ emission (ZPL wavelength at 575\,nm), our measured brightness underestimates the actual efficiency. For an ideal characterization, the brightness can be improved by including a charge initialization sequence in the measurement. Finally, photon purity can be improved with a more efficient delivery of the excitation laser, along with optimized spectral and spatial filtering.   

We have described the design, fabrication, and evaluation of an all-diamond parabolic reflector located around a single NV center. Already in this proof-of-principle demonstration, to the best of our knowledge, our results represent the highest photon flux from a single NV center and also the highest brightness of any quantum optical device in diamond. With the current system performance, detection of multiple photons at the microsecond time scale is already possible, opening exciting new prospects in single-shot high-fidelity spin readout~\cite{steiner2010universal,gupta2016efficient}. As this device incorporates implanted NV centers at a depth of 100\,nm, it should also be possible to observe near lifetime-limited linewidths \cite{chu2014coherent} for high rate, high visibility quantum interference, which will be investigated at low temperatures in a future work. With such an NV center in a parabolic reflector, the present extraction efficiency should lead to an order-of-magnitude improvement in photon-mediated electron-electron entanglement generation rates~\cite{bernien2013heralded,pfaff2014unconditional,hensen2015loophole,kalb2017entanglement}. When integrated with a near-surface NV center, it is also possible to realize scanning NV magnetometers for exploring exotic condensed matter systems~\cite{balasubramanian2008nanoscale,maletinsky2012robust,rondin2013stray,appel2016fabrication,pelliccione2016scanned}. Although demonstrated with NV centers in diamond, this robust, spectrally broadband platform could be directly integrated with other emerging color centers such as the silicon-vacancy and germanium-vacancy centers in diamond, or for other quantum emitters in numerous materials~\cite{aharonovich2016solid}.

\section*{Methods}
\subsection*{Sample preparation}

The sample consisted of a 50-$\mu$m-thick high-purity type IIA CVD-grown diamond sample (Element6) implanted with $^{15}$N at an energy of 85 keV (depth $\sim 100$\,nm) at a dosage of $10^9$ $^{15}$N/cm$^2$. We subsequently annealed the sample at 1200 $\degree$C to form NV centers following the recipe detailed in Ref.~\cite{chu2014coherent}. Finally, we cleaned the diamond in a boiling mixture of 1:1:1 sulphuric, nitric and perchloric acid. 

\subsection*{Device fabrication}
We deposited 200\,nm-thick silicon nitride (SiN) via plasma-enhanced chemical vapor deposition. Electron-beam lithography defined an annulus in a positive-tone electron-beam resist (ZEP\,520A exposed at 500\,$\mu$C/cm$^2$ and developed at 0$^{\circ}$C in ortho-xylene for 90\,s) around a pre-registered NV center. The inner diameter of the annulus determined the device diameter, and the developed area exposed the material to dry etching. Following development, we thermally reflowed the resist above its glass transition temperature at an optimized temperature of 215 $\degree$C for 15 minutes. We used a tetrafluoromethane (CF$_4$) plasma to transfer this pattern into SiN. Finally, we produced parabolic structures in diamond by transferring the SiN pattern using inductively coupled oxygen plasma reactive-ion etching (0.15\,Pa, bias power 500\,W, RF power 240\,W). We characterized the device using atomic-force microscopy (AFM) to illustrate the principles that enable the production of parabolic structures.  In Fig.\,\ref{fab}(b), the AFM is performed on a different diamond chip to prevent contamination of the high-purity diamond sample used in the experiment. The fabrication parameters are identical to those in the main text and above, except for the SiN thickness (125nm) and diamond etch depth. Because the AFM tip is asymmetrical, we show the cross-section for only one side of the device. The shaded area is a fit to a spherical surface in steps\,(ii)-(iii) and parabolic surface in step\,(iv). The final RMSE of 21.23\,nm indicates a good fit and smooth surface down to $\lambda/12n$ where $\lambda$ is the NV$^{-}$ ZPL wavelength. The flat surface in step (iv) is due to premature etch termination. 

\subsection*{Experiment}

We collected and detected the fluorescence using a 1.3 NA $\times$100 oil-immersion infinity-corrected objective (Nikon Plan Fluor)and a single photon avalanche diode (SPAD, Excelitas SPCM-AQRH). The quantum efficiency of the SPAD is $\eta_{1}\approx 0.6868$ for a typical NV spectrum. The setup has a transmission of $\eta_{2}\approx 0.3696$. The total setup efficiency is $\eta_{1} \eta_{2} \approx 0.2538$. When excited at 532 nm (Coherent, Sapphire 532 SF; NKT Photonics, SuperK Extreme with tunable filter), the maximum fluorescence detection rate $F$ increases with increasing power in accordance with Eq.(\ref{brightness})-(\ref{excitation_prob}). As such it is conventional to represent $F$ by its asymptotic limit as $P\rightarrow \infty$. This limit can be obtained by fitting the measured count rates to Eq.(\ref{brightness}). 

\bibliography{reflectorref2017}

\begin{acknowledgments}
We are grateful to Stephan G{\"o}tzinger, Karl Berggren, James Daley, Mark Mondol, and Tim Savas for helpful discussions. This research was supported in part by the Army Research Laboratory Center for Distributed Quantum Information (CDQI). N.H.W was supported by CDQI. D.K. was partially covered by the Kwanjeong scholarship from the Kwanjeong educational foundation. S.M. was supported by the NSF program ACQUIRE: Scalable Quantum Communications with Error-Corrected Semiconductor Qubits. B.L and M.W. were supported in part by CDQI and the STC Center for Integrated Quantum Materials (CIQM), NSF Grant No. DMR-1231319. We acknowledge help with AFM by Tsung-Ju Lu. 
\end{acknowledgments}

\end{document}